\documentclass[showpacs,twocolumn,pra,prd,superscriptaddress,longbibliography,notitlepage]{revtex4-2}
\usepackage[dvips]{graphicx}
\usepackage{amsmath,amssymb,amsthm,mathrsfs,amsfonts,dsfont}
\usepackage{subfigure, epsfig}
\usepackage{braket}
\usepackage{bm}
\usepackage{bbm}
\usepackage{enumerate}
\usepackage{color}
\usepackage{comment}
\usepackage{diagbox}
\usepackage{youngtab}
\usepackage{lineno}
\usepackage{upgreek}
\usepackage[colorlinks = true]{hyperref}

\graphicspath{{./figure/}}

\newtheorem{observation}{Observation}

\newtheorem{lemma}{Lemma}

\newtheorem{definition}{Definition}
\newtheorem{proposition}{Proposition}

\newcommand{\tr}{\mathrm{Tr}}

\newcommand{\red}[1]{{\color{red} #1}}

\newcommand{\comments}[1]{}
\newcommand{\sw}{\sqrt{\mathrm{SWAP}}}

\begin{document}
\title{A scheme to create and verify scalable entanglement in optical lattice}

\author{You Zhou}
\altaffiliation{These authors contributed equally to this work}
 \affiliation{Department of Modern Physics,
 University of Science and Technology of China, Hefei, Anhui 230026, China}
\affiliation{Key Laboratory for Information Science of Electromagnetic Waves (Ministry of Education), Fudan University, Shanghai 200433, China}

\author{Bo Xiao}
\altaffiliation{These authors contributed equally to this work}
 \affiliation{Department of Modern Physics,
 University of Science and Technology of China, Hefei, Anhui 230026, China}
 \affiliation{
	CAS Center for Excellence and Synergetic Innovation Center in Quantum Information and Quantum Physics,
	University of Science and Technology of China, Hefei, Anhui 230026, China
}%

\author{Meng-Da Li}%
 \affiliation{Department of Modern Physics,
 University of Science and Technology of China, Hefei, Anhui 230026, China}
 \affiliation{
	CAS Center for Excellence and Synergetic Innovation Center in Quantum Information and Quantum Physics,
	University of Science and Technology of China, Hefei, Anhui 230026, China
}%

\author{Qi Zhao}
\affiliation{Joint Center for Quantum Information and Computer Science, University of Maryland, College Park, Maryland 20742, USA}

\author{Zhen-Sheng Yuan}
\email{yuanzs@ustc.edu.cn}
\affiliation{Department of Modern Physics,
 University of Science and Technology of China, Hefei, Anhui 230026, China}
 \affiliation{
	CAS Center for Excellence and Synergetic Innovation Center in Quantum Information and Quantum Physics,
	University of Science and Technology of China, Hefei, Anhui 230026, China
}%
\author{Xiongfeng Ma}
\email{xma@tsinghua.edu.cn}
 \affiliation{
Center for Quantum Information, Institute for Interdisciplinary Information Sciences, Tsinghua University, Beijing 100084, China
}%

\author{Jian-Wei Pan}
\affiliation{Department of Modern Physics,
 University of Science and Technology of China, Hefei, Anhui 230026, China}
 \affiliation{
	CAS Center for Excellence and Synergetic Innovation Center in Quantum Information and Quantum Physics,
	University of Science and Technology of China, Hefei, Anhui 230026, China
}%

\comments{
\begin{abstract}
To achieve scalable quantum information processing, great efforts have been devoted to the creation of large-scale entangled states in various physical systems. Ultracold atom in optical lattice is considered as one of the promising platforms
due to its feasible initialization and parallel manipulation. In this work, we propose an efficient scheme to generate and characterize global entanglement in the optical lattice. With only two-layer quantum circuits, the generation utilizes two-qubit entangling gates based on the superexchange interaction in double wells. The parallelism of these operations enables the generation to be fast and scalable. To verify the entanglement of this non-stabilizer state, we mainly design three complementary detection protocols which are less resource-consuming compared to the full tomography. In particular, one just needs two homogenous local measurement settings to identify the entanglement property. 
Our entanglement generation and verification protocols provide the foundation for the further quantum information processing in optical lattice.
\end{abstract}
}

 \date{\today}

\maketitle

\section{Abstract}
To achieve scalable quantum information processing, great efforts have been devoted to the creation of large-scale entangled states in various physical systems. Ultracold atom in optical lattice is considered as one of the promising platforms due to its feasible initialization and parallel manipulation. In this work, we propose an efficient scheme to generate and characterize global entanglement in the optical lattice. With only two-layer quantum circuits, the generation utilizes two-qubit entangling gates based on the superexchange interaction in double wells. The parallelism of these operations enables the generation to be fast and scalable. To verify the entanglement of this non-stabilizer state, we mainly design three complementary detection protocols which are less resource-consuming compared to the full tomography. In particular, one just needs two homogenous local measurement settings to identify the entanglement property. Our entanglement generation and verification protocols provide the foundation for the further quantum information processing in optical lattice.

\section{Introduction}
Quantum information and quantum computation \cite{Nielsen2011Quantum}, which harvest the intrinsic quantum features, like superposition and entanglement \cite{Horodecki2009entanglement}, can show advantages against their classical counterparts. To build a practical quantum information processor, the computation platform with high scalability is preferred and the qubits should be coupled to each other to form a large-scale entanglement. Hence, many researches have been focusing on the generation of scalable entangled states in different physical systems, e.g, ion trap \cite{Monz2011,Nigg2014ion,Friis2018}, photons \cite{Wang2018,Zhong2018}, Rydberg atoms \cite{Omran2019} and superconducting circuits \cite{Gong2019,Wei2020,Song2019}.
Even though there are significant progresses of the qubit number in various systems \cite{Zhang2017dynamical,Arute2019Supremacy}, the generation of large-scale entanglement is still challenging for the Noisy Intermediate Scale Quantum
Devices \cite{Preskill2018NISQ}.

Ultracold atoms in optical lattice \cite{Bloch2012ultracold} could be a practicable system to overcome this challenge due to its feasible initialization and parallel manipulation.
By adiabatically increasing the lattice depth, the phase of ultracold atom can be tuned from superfluid (SF) to Mott insulator (MI) \cite{Jaksch1998,Greiner2002}. Under an unit filling rate, numerous atoms can be confined in the lattice and serves as qubits. Based on this initialization, entangled states in optical lattice have been demonstrated experimentally,
for instance, the generation of cluster state with controlled collision gate induced by the spin-dependent lattice \cite{Jaksch1999, Mandel2003}, which acts as the resource state for the measurement-based quantum computing \cite{onewayQC,Raussendorf2003Measurement}. The development of superlattice further improves the ability to control ultracold atoms, which is formed by overlapping two different optical lattices to generate a series of double wells. The structure of the double well can be modified to induce different kinds of atomic dynamics, such as superexchange coupling \cite{Trotzky2008} and controlled exchange interaction \cite{Anderlini2007}, which can be both used to realize $\sqrt{\mathrm{SWAP}}$ gate and entangle two atoms in a double well \cite{Dai2016,Anderlini2007}. 
Besides, the entangling operations in the superlattice can be performed in a parallel way based on the periodicity of the lattice system, which is suitable for the fast generation of large amount of entangled pairs and even large-scale entangled states. Note that this kind of parallel operation can also be implemented with multi-tweezer in Rydberg atom experiments \cite{Scholl2021,PScholl2}.

Every coin has two sides. The periodicity property also induces some restrictions on the quantum operation and measurement. The small lattice spacing, required by the large tunnelling between neighbour sites, creates a challenge for the individual control of atoms, say, some local basis rotations. The tight-focused optical tweezer \cite{Weitenberg2011} created with the recently developed high-resolution imaging \cite{Bakr2009,Sherson2010,Parsons2015,Cheuk2015,Haller2015} could be a solution. However, it is still challenging to perform a few of different single-qubit operations on multiple qubits under a realistic time-scale to ensure the system coherence. As a result, homogeneous operations and measurements on all atoms are preferred in optical lattice experiments.

In this work, we propose a scheme to generate scalable entanglement of ultracold atoms which is suitable for the implementation in the optical superlattice system. It mainly contains two entangling steps: first, entangle the atom-pair in each double well by the $\sqrt{\mathrm{SWAP}}$ gate; second, shift the position of double wells to a single site by changing the phase of superlattice, and then entangle the the new atom-pair with $\sqrt{\mathrm{SWAP}}$ again.
In this way, all atoms in the superlattice can be connected with neighbours to form a global entangled state. Our theoretical analysis shows that the final state possesses genuine multipartite entanglement (GME). In addition, the final state is also less sensitive to magnetic noises which can cause decoherence, since it only owns amplitude on the computational basis whose total spin is zero.

In actual experiments, the inevitable noise may degrade the entanglement, which should be verified further. Compared with quantum tomography \cite{Vogel1989Determination,Paris2004esimation}, entanglement witness is a more efficient way \cite{TERHAL2001witness,GUHNE2009detection,Friis2019Reviews} to realize this task based on the pre-knowledge of the preparation. Current entanglement witnesses are usually designed for some structured states, such as permutation-invariant states \cite{Toth2009symmetric,Toth2010Permutation,Zhou2019Decomposition} and stabilizer states \cite{Toth2005Detecting,Knips2016Multipartite,You2019graph}. However, in our protocol the state generated by the non-Clifford gate $\sw$ is not a standard stabilizer state, which makes the entanglement verification challenging. To overcome this challenge, we first construct an entanglement witness based on the preparation fidelity. By adopting the decomposition using stabilizer-like method, we can lower bound the fidelity with a few spin-correlation measurements. Some correlations among them are inhomogenous and thus require the individual atom addressing whose implementation could be challenging with current techniques. To further ease the experiment realization, we show another complementary protocol which only requires homogenous spin-correlations and only two measurement settings. At last, by reversely evolving the state with the conjugate quantum gates, we provide an intuitive method to indirectly bound the preparation fidelity and thus qualitatively verify the entanglement.

\section{Results}
\subsection{Entangling Gates in Optical Superlattices}

\begin{figure}[htbp]
  \centering
  \includegraphics[width=1\linewidth]{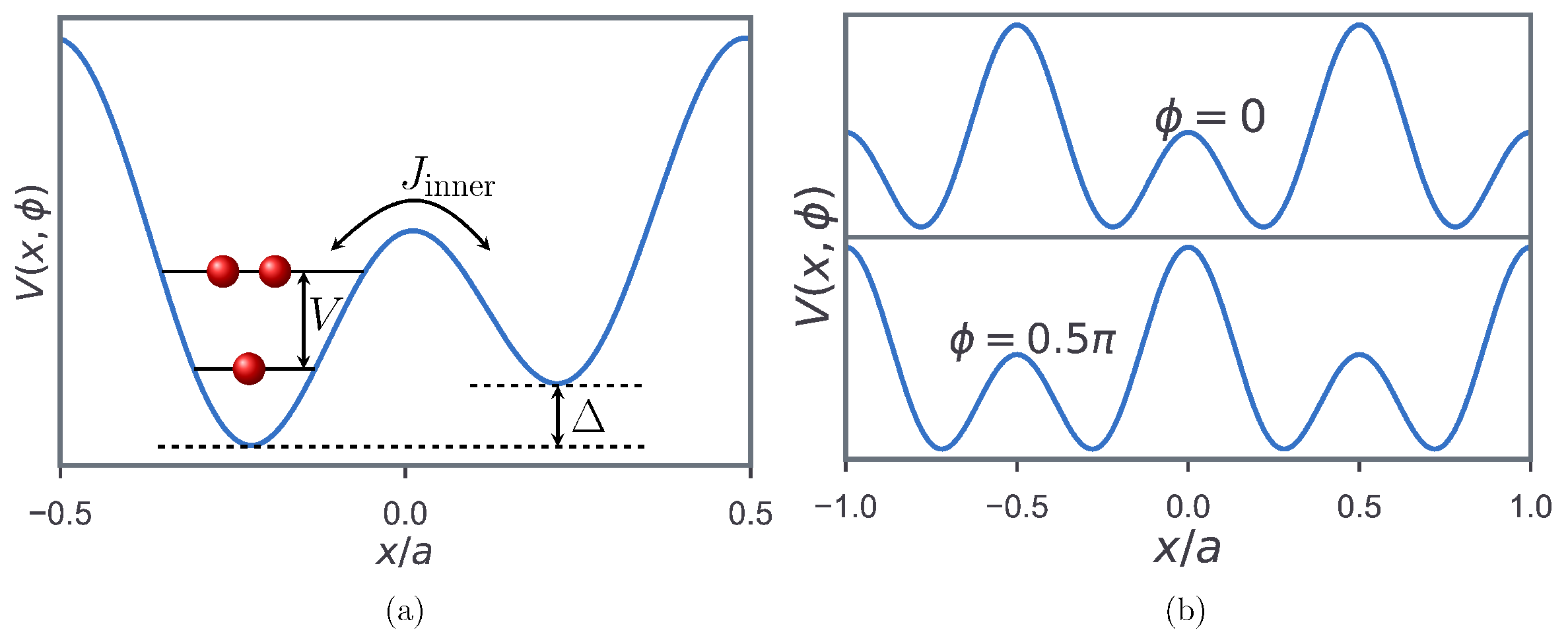}
  \caption{The features of superlattice. (a) Double-well system characterized by the parameters $V$, $J_{\mathrm{inner}}$ and $\Delta$. (b) The shift of the double-well center.}\label{Fig1}
\end{figure}

The behavior of ultracold atoms in optical lattices can be described by the Hubbard model which is characterized by the tunnel coupling between neighbouring sites $J$, the on-site interaction $V$, and the effective chemical potential $\mu$. By increasing the depth of the lattice, a phase transition from SF to MI can occur and the atoms start to be localized in each site. To extend the ability to manipulate atoms, a more complicated periodic potential, named superlattice, was proposed and applied in many experiments \cite{SebbyStrabley2006,Lee2007,Trotzky2010,Lohse2015}. It is normally formed by overlapping two distinct optical lattices. The period of the first lattice (denoted as the long lattice) is twice as that of the second (denoted as the short lattice), which induces an array of double wells steadily.  The resulting potential shows $V_{\mathrm{t}}(x,\phi) = V_{\mathrm{l}}\text{cos}^2(\uppi x/a+\phi) +V_{\mathrm{s}}\text{cos}^2(2\uppi x/a)$ with $a$ being the lattice spacing of the long lattice. The structure of such potential is dependent on both the relative strength of two lattices and the relative phase $\phi$. When the phase $\phi$ is not equal to $n\uppi/2$ ($n$ is an integer), all the double wells are biased with a non-zero tilt $\Delta$ between the subsites. In addition, the center of each double well can shift a period of short lattice via changing n by an odd number, which is illustrated in Figure \ref{Fig1} (b).

\onecolumngrid

\twocolumngrid

Subsites in each double well are separated by a low barrier, while neighbour double wells are separated by a high barrier. We denote corresponding tunneling strengths as $J_{\mathrm{inter}}$ and $J_{\mathrm{inner}}$ respectively. As $J_{\mathrm{inter}}$ is much smaller than $J_{\mathrm{inner}}$, the hopping event between different double wells can be ignored and the movement of atoms is restricted in each double well. In this case, for two-level bosonic atoms, their dynamics in double wells can be described well with a two-site mode Bose-Hubbard Hamiltonian characterized with $V$, $J_{\mathrm{inner}}$ and $\Delta$ as follows.

\begin{equation}\label{Eq:bhmodel}
  \begin{aligned}
    \hat{H}=&-J_{\mathrm{inner}}\sum_{\sigma=\uparrow,\downarrow}(\hat{a}_{\mathrm{L},\sigma}^{\dag}\hat{a}_{\mathrm{R},\sigma}+\hat{a}_{\mathrm{R},\sigma}^{\dag}\hat{a}_{\mathrm{L},\sigma})\\
    &+\frac{V}{2}\sum_{i=\mathrm{R},\mathrm{L}}
    \hat{n}_{i,\uparrow}(\hat{n}_{i,\uparrow}-1) +\frac{V}{2}\sum_{i=\mathrm{R},\mathrm{L}}
    \hat{n}_{i,\downarrow}(\hat{n}_{i,\downarrow}-1)\\
    &+V\sum_{i=\mathrm{R},\mathrm{L}}(\hat{n}_{i,\uparrow}\hat{n}_{i,\downarrow}+\hat{n}_{i,\downarrow}\hat{n}_{i,\uparrow})
    +\frac{\Delta}{2}(\hat{n}_{\mathrm{L},\uparrow}-\hat{n}_{\mathrm{R},\uparrow}-\hat{n}_{\mathrm{L},\downarrow}+\hat{n}_{\mathrm{R},\downarrow}).
  \end{aligned}
\end{equation}

Here, $\mathrm{R}(L)$ denotes right(left) subsite, $\hat{a}_{i,\sigma}^{\dag}$($\hat{a}_{i,\sigma}$) is the creation(annihilation) operator for the boson on site i with the inner state $\sigma$, and $\hat{n}_{i,\sigma}$ is the number operator. We consider a non-biased double-well system starting with unit filling---one atom per site. Note that the chemical potential $\mu$ is fixed here for unit filling and thus has no effect on the dynamics, and it is ignored in Eq.~\eqref{Eq:bhmodel}. In the limit $V \gg J_{\mathrm{inner}}$, the large $V$ prevents multiple occupation, so that the system would evolve in a subspace consists of four basis states labeled by inner states of the bosons $\ket{\uparrow,\uparrow}$, $\ket{\downarrow,\uparrow}$, $\ket{\uparrow,\downarrow}$ and $\ket{\downarrow,\downarrow}$. By using perturbation theory, the above model in this subspace is equivalent to the isotropic Heisenberg Hamiltonian \cite{Duan2003}:

\begin{equation}
    \hat{H}_{\mathrm{eff}}=-\frac{J_{\mathrm{ex}}}{2}(\hat{X}_{\mathrm{L}}\hat{X}_{\mathrm{R}}+\hat{Y}_{\mathrm{L}}\hat{Y}_{\mathrm{R}}+\hat{Z}_{\mathrm{L}}\hat{Z}_{\mathrm{R}}),
\end{equation}
where $\hat{X}_{i}$, $\hat{Y}_{i}$, $\hat{Z}_{i}$ are Pauli operators on subsite $i=\mathrm{R}, \mathrm{L}$, and $J_{\mathrm{ex}}\approx 2J_{\mathrm{inner}}^2/V$ is the superexchange coupling between subsites.

In the remaining of this work, we denote spin configuration $\ket{\downarrow}$ and $\ket{\uparrow}$ as $\ket{0}$ and $\ket{1}$. Initialized at $\ket{0,1}$ and driven by this effective Hamiltonian, the state of the system can oscillate between $\ket{0,1}$ and $\ket{1,0}$ with a period of $T=2\uppi\hbar/J_{\mathrm{ex}}$ while global phase is also recovered, named the superexchange process. Taking an evolution time of $T/8$, the product states $\ket{0,1}$ and $\ket{1,0}$ are prepared into two-qubit maximally entangled states $1/\sqrt{2}(\ket{0,1}+i\ket{1,0})$ and $1/\sqrt{2}(\ket{0,1}-i\ket{1,0})$ respectively while $\ket{0,0}$ and $\ket{1,1}$ remain unchanged due to the high energy gap of $V$. The effective unitary transformation corresponds to a $\sqrt{\mathrm{SWAP}}^{\dagger}$ gate operation, i.e.,

\begin{equation}
    U_{\sqrt{\mathrm{SWAP}}^{\dagger}}=
    \begin{pmatrix}
    1 & 0 & 0 &0 \\ 
    0 & (1-i)/2 & (1+i)/2 & 0\\ 
    0 & (1+i)/2 & (1-i)/2 & 0 \\ 
    0 & 0 & 0 & 1
    \end{pmatrix}.
\end{equation}
with respect to the basis $\ket{0,0}$, $\ket{1,0}$, $\ket{0,1}$, $\ket{1,1}$. Moreover, with an evolution time of $3T/8$, one can realize the corresponding $\sqrt{\mathrm{SWAP}}$ gate. In particular, when $J_{\mathrm{inter}}/J_{\mathrm{ex}}\geq 25$, the infidelity of the $\sqrt{\mathrm{SWAP}}^{\dagger}$ operation caused by the tunneling between neighbour double wells would be smaller than $0.1\%$.

For the current gate generation scheme, due to the large ratio $V/J_{\mathrm{inner}}$, the effective $J_{\mathrm{ex}}$ is far less than $J_{\mathrm{inner}}$. In addition, such high $V/J_{\mathrm{inner}}$ requires a deep potential in each site thus leads to a small $J_{\mathrm{inner}}$. As a result, the period of the superexchange process would be long, e.g, tens of milliseconds for $^{87}$Rb atoms in superlattice \cite{Trotzky2008,Dai2016}, which would aggravate the decoherence effect. To ease this problem, an alternative and faster gate generation scheme can be adopted here \cite{Yang2020}, which is performed with a small $V/J_{\mathrm{inner}}$. Instead of introducing the large energy ratio, this scheme utilizes the coherent competition of superexchange and atom tunneling to decrease the component with double occupation in state. As $V/J_{\mathrm{inner}}$ is set to be a finite value $4/\sqrt{3}$, the undesired component can be eliminated completely at specific time intervals. In particular, with an evolution time of $\uppi\hbar/V$, such elimination leads to a fast $\sqrt{\mathrm{SWAP}}^{\dagger}$ gate realization with high fidelity.

Besides, the spin-dependent effect further improves the ability to manipulate the atoms in optical lattices. By adding circular polarization components into one lattice light field of superlattice \cite{Lee2007,Yang2017}, the tilt $\Delta$ can be different for different inner states. Therefore, the energy gap between inner states on right subsite would be different from that on the left which induce two applications here. One is to spin flip atoms in one subsite of every double well without affecting atoms in the other subsite, enabling the state transfer between the four basis states mentioned above \cite{Lee2007}. The other is to generate a relative phase between $\ket{0,1}$ and $\ket{1,0}$ and then transfer the states $1/\sqrt{2}(\ket{0,1}-i\ket{1,0})$, $1/\sqrt{2}(\ket{1,0}-i\ket{0,1})$ to the standard Bell states $1/\sqrt{2}(\ket{1,0}+\ket{0,1})$ , $1/\sqrt{2}(\ket{1,0}-\ket{0,1})$ \cite{Dai2016}, respectively. 

\subsection{Entanglement generation protocol}\label{subsec:Scheme}

\begin{figure}[htbp]
 \centering
 \includegraphics[width=0.9\linewidth]{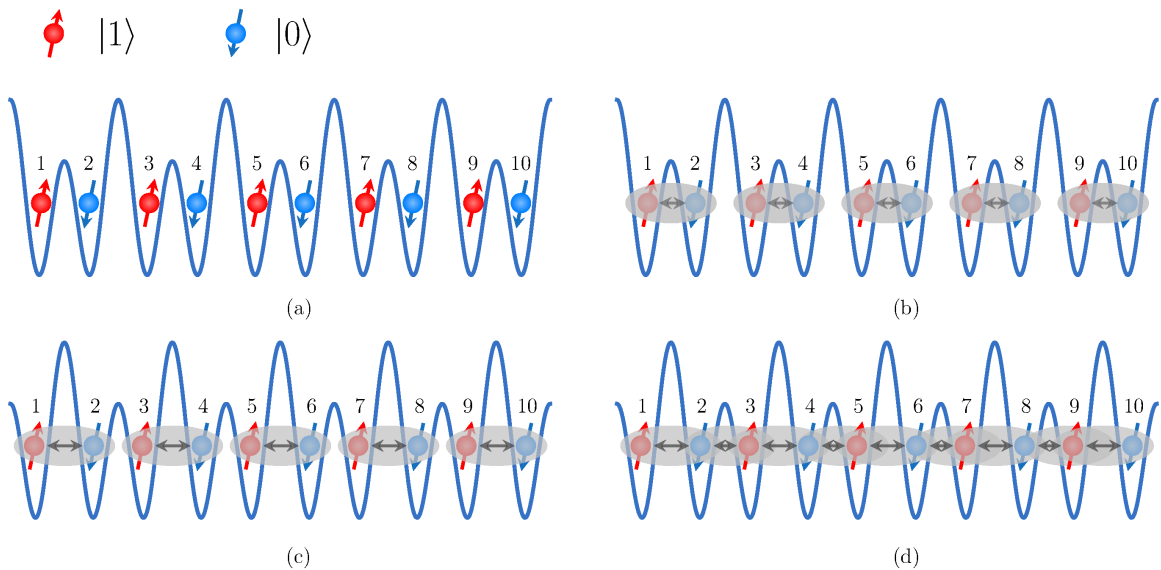}
  \caption{Illustration of the preparation protocol. (a) Generate Néel state with selective spin-flip. (b) Entangle and prepare the atom pairs into Bell state with $\sqrt{\mathrm{SWAP}}^{\dagger}$ gate. (c) Shift the position of double wells by changing the relative phase $\phi$. (d) Entangle the neighbour atoms which are not entangled in (b).}
  \label{fig:thescheme}
\end{figure}

Our protocol is expected to be performed on one-dimensional atomic chains along X direction with a short lattice isolating the atoms. On each site, the filling is initialized into unit through the SF-MI phase transition. Assuming the inner state of atoms are prepared into $\ket{0}$ and taking 10-qubit system as example, the protocol can be divided into the following steps which are illustrated in Fig.~\ref{fig:thescheme}.

\begin{enumerate}
  \item Turn on the spin-dependent superlattice \cite{Yang2017} by ramping up a spin-dependent long lattice along X. With selective spin-flip by coupled field, the Néel state is prepared:
  \begin{equation}\label{neel}
      \ket{\Phi_{1}}=\ket{1010101010}
  \end{equation}
  \item Turn off the spin-dependent effect and perform $\sqrt{\mathrm{SWAP}}^{\dagger}$ gate on the atom pairs $(1,2), (3,4),\cdots,(9,10)$ in each double well,
  \begin{equation}
\begin{aligned}
\ket{\Phi_2'}=&\bigotimes_{k=1}^5\sqrt{\mathrm{SWAP}}^\dag_{\{2k-1,2k\}} \ket{\Phi_1}\\
=&\bigotimes_{k=1}^5\frac1{\sqrt{2}}(\ket{10}+i\ket{01})_{\{2k-1,2k\}}.
\end{aligned}
\end{equation}
  \item Turn on the spin-dependent effect again and introduce a relative phase $\uppi/2$ on $\ket{10}$. As a result, the state of the system is the tensor product of five Bell states,
  \begin{equation}
\begin{aligned}
\ket{\Phi_2}=U_{\mathrm{phase}}\ket{\Phi_2'}=\bigotimes_{k=1}^5\frac1{\sqrt{2}}(\ket{01}+\ket{10})_{\{2k-1,2k\}}.
\end{aligned}
\end{equation}
 \item Change the relative phase $\phi$ by $\uppi/2$ to trap the atom pair $(2,3), (4,5),\cdots,(8,9)$ in different double wells.
 \item Perform $\sqrt{\mathrm{SWAP}}^{\dagger}$ gates on the atom pairs $(2,3), (4,5),\cdots,(8,9)$ in each double well, the final state is
     \begin{equation}
    \begin{aligned}
    \ket{\Psi}=\bigotimes_{k=1}^4\sqrt{\mathrm{SWAP}}^\dag_{\{2k,2k+1\}} \ket{\Phi_2}.
    \end{aligned}
    \end{equation}
\end{enumerate}

\onecolumngrid

\twocolumngrid

During the entire process atomic motions along other directions are frozen by deep lattice potentials. Finally, one-dimensional entangled systems are generated. The gate operations entangle every pair of neighbour atoms so that the target state is GME by itself. As the gate operation is not present, the depth of each site should be deep enough to avoid any undesirable hopping events.

The target state only owns non-zero projection on a few bases whose total spin numbers are zero, and the magnetic field fluctuation can only introduce a global phase. Consequently, our target can be robust to the magnetic field fluctuation, and thus has longer coherence time, compared to for example the GHZ state. In this basis collection, each basis can find another basis whose spin on each qubit is anti-parallel to its own, e.g. $\ket{0,1,1,0,...}$ and $\ket{1,0,0,1,...}$, and the projections on each pair of bases show same modules but different arguments.

According to the definition of GME \cite{Horodecki2009entanglement}, it is inferred that a pure state whose subsystems are all mixed states must be a GME state. Based on this, the purity of all subsystems of the target state in the case of 10-qubit are calculated and found to be less than 1, which verifies that this 10-qubit target state possesses GME. This conclusion can be extended to the target state with a different number of qubits. 

We remark that Ref.~\cite{Vaucher2008} also studies the generation of entanglement by $\sqrt{\mathrm{SWAP}}$ in the cold atom system. The state generated there is like a GHZ state, which needs much more gate operations, compared to the cluster-like state here. In addition, there is no valid method given there to verify the entanglement.

In the following sections, we exhibit theoretical methods to detect the multipartite entanglement in the cold atom system. The prepared state shown in the Sec.~\ref{subsec:Scheme} is generated by parallel $\sqrt{\mathrm{SWAP}}^\dag$ gates and thus not a stabilizer state, such as the GHZ state and the cluster state. This fact makes the entanglement detection task challenging and the methods proposed in literature are not suitable in this scenario. 

To overcome this challenge, we propose three complementary methods. The first one is based on the fidelity estimation to the target state to detect the strongest form of entanglement---genuine multipartite entanglement. By evaluating the non-Pauli stabilizer after the entangling gate evolution, the method only needs constant number of measurement setting with respect to the system size to lower bound the fidelity. To further release the experiment efforts, the second method adopts homogeneous measurements, with only two measurement settings. Thus it is very efficient to realize and can tell whether the prepared state is separable or not considering any bipartition of the whole system. The third one is more intuitional but can reveal GME with less experiment requirements. In particular, we indirectly estimate the fidelity between the prepared state and the target state by evolving the entangling process forward and backward. 

All these methods need the spin measurement on single atom, however the fluorescence imaging used in high-resolution experiments can not resolve different spins. As the state we consider here only has one atom on each site, it could be done by measuring the atom distribution after removing one spin component in which the occupied site and unoccupied site represent different spins \cite{doi:10.1126/science.aag1430} or splitting different spin components with gradient along perpendicular direction \cite{doi:10.1126/science.aag1635}.

Before showing these three methods, we give some related definitions about multipartite entanglement.

A pure state is (bi-)separable if it is in a tensor product form $\ket{\Psi_b}=\ket{\Phi_{A}}\otimes \ket{\Phi_{\bar{A}}}$, where $\mathcal{P}_2=\{A, \bar{A}\}$ is a bipartition of the qubits in the system. A mixed state is separable if it can be written as a mixing of pure separable states,
$\rho_\mathrm{b}=\sum_ip_i\ket{\Psi_\mathrm{b}^i}\bra{\Psi_\mathrm{b}^i}$. Note that each separable state $\ket{\Psi_\mathrm{b}^i}$ in the summation can have different bipartitions. The separable state set is denoted as $S_\mathrm{b}$. There is another restricted way for the extension to mixed states. A state is $\mathcal{P}_2$-separable, if it is a mixing of pure separable states with a same partition $\mathcal{P}_2$, and we denote the state set as $S_\mathrm{b}^{\mathcal{P}_2}$. It is clear that $S_\mathrm{b}^{\mathcal{P}_2}\subset S_\mathrm{b}$, and $S_\mathrm{b}$ can be generated by the convex mixture of all possible $S_\mathrm{b}^{\mathcal{P}_2}$.

\begin{definition}\label{Def:fully}
An $N$-qubit quantum state $\rho$ is fully entangled, if it is outside of the separable state set $S_\mathrm{b}^{\mathcal{P}_2}$ for any bipartition,
\begin{equation}\label{Eq:fully}
\rho \notin S_\mathrm{b}^{\mathcal{P}_2}, \forall \mathcal{P}_2=\{A, \bar{A}\}.
\end{equation}
\end{definition}

\begin{definition}\label{Def:GME}
An $N$-qubit quantum state $\rho$ possesses genuine multipartite entanglement, if it is outside of the separable state set $S_\mathrm{b}$,
\begin{equation}\label{Eq:GME}
\rho \notin S_\mathrm{b}.
\end{equation}
\end{definition}
Since $S_\mathrm{b}^{\mathcal{P}_2}\subset S_\mathrm{b}$, GME is a stronger claim than full entanglement. We also remark that the recently demonstrated entanglement in the IBM cloud quantum computing \cite{Wang2018IBM} is actually the full entanglement defined here. By Def.~\ref{Def:fully}, for a state with full entanglement, it is possible to prepare it by mixing bi-separable states with different bipartitions \cite{Guehne2009}. On the other hand, GME describes the strongest form of quantum entanglement, that is, all the qubits in the system are indeed entangled with each other. GME is essential in various multipartite quantum information tasks, such as quantum cryptography \cite{Chen2007Multi}, quantum nonlocality \cite{brunner2014bell}, quantum networks \cite{wehner2018quantum,perseguers2013distribution}, quantum metrology \cite{giovannetti2011advances} and measurement-based quantum computing \cite{Raussendorf2001One}.

\subsection{Entanglement detection based on fidelity estimation}\label{Sec:fidelity}
In this section, we show an entanglement detection protocol based on the fidelity value between the prepared state and the target state.
\begin{proposition} \label{Th:Fwitness}
The operator $\mathcal{W}_{\Psi}$ can witness genuine multipartite entanglement near $\ket{\Psi}$,
\begin{equation}\label{Eq:Fwitness}
\mathcal{W}_{\Psi}=\frac{5}{8}\mathbb{I}-\ket{\Psi}\bra{\Psi},
\end{equation}
with $\langle \mathcal{W}_{\Psi} \rangle \ge0$ for any separable state in $S_\mathrm{b}$.
\end{proposition}

According to Proposition \ref{Th:Fwitness}, if the fidelity of the prepared state $\rho_\mathrm{pre}$ with the target state $\ket{\Psi}$, i.e., $\mathrm{Tr}(\rho_\mathrm{pre}\ket{\Psi}\bra{\Psi})$, exceeds $\frac{5}{8}$, $\rho_\mathrm{pre}$ possesses GME. However, it is generally difficult to evaluate the quantity $\mathrm{Tr}(\rho_\mathrm{pre}\ket{\Psi}\bra{\Psi})$ by the direct projection on $\ket{\Psi}$, as it is an entangled state.

Alternatively, one needs to decompose the density operator $\Psi$ into the summation of many local measurements in the form $\otimes_{i=1}^{N}O_i$, which is easier to implement in experiments. Here $O_i$ is Hermitian operator of the i-th qubit. The number of local measurements characterizes the experiment effort to estimate the fidelity.

Here, in order to reduce the measurement effort, instead of direct decomposing $\Psi$, we give a lower bound of the fidelity by using the stabilizer-like operator for the non-stabilizer state $\Psi$.
\begin{proposition}\label{Th:lowerbound}
For the target state $\ket{\Psi}\bra{\Psi}$ and its $N$ independent stabilizers $S_i'$, the following inequality holds,
\begin{equation}\label{Eq:OpIneq}
\ket{\Psi}\bra{\Psi}\ge\frac1{2}\sum_{i=1}^N S_i'-\left(\frac{N}{2}-1\right)\mathbb{I},
\end{equation}
where $A\geq B$ indicates that $(A-B)$ is positive semidefinite. The stabilizer is determined by the evolution of $\sw$ gates $S_i'=US_iU^{\dag}$ with $U=\bigotimes_{k=1}^{\frac{N}{2}-1}\sqrt{\mathrm{SWAP}}^\dag_{\{2k,2k+1\}}$, where
\begin{equation}\label{Eq:StaBell}
\begin{aligned}
S_{2k-1}=X_kX_{k+1}, S_{2k}=-Z_kZ_{k+1},
\end{aligned}
\end{equation}
for $k=1,\cdots N/2$, are the stabilizers for the Bell pairs.
\end{proposition}

Due to $\sw^{\dag}$ is not a Clifford gate, the corresponding stabilizer $S_i'$ is not in a tensor product of Pauli operators, but the summation of them. One can directly get $S_i'$ by the evolution of the parallel $\sw^\dag$ on $S_i$. Due to the locality of the evolution, $S_i$ can only be transformed by the gate with overlapping support. For example,
\begin{equation}
  \begin{split}
S_1'=U_{2,3}X_{1}X_{2}U^{\dagger}_{2,3}&=\frac{1}{2}X_{1}(X_{2}I_{3}+I_{2}X_{3}-Y_{2}Z_{3}+Z_{2}Y_{3}),\\
S_2'=-U_{2,3}Z_{1}Z_{2}U^{\dagger}_{2,3}&=-\frac{1}{2}Z_{1}(Z_{2}I_{3}+I_{2}Z_{3}+Y_{2}X_{3}-X_{2}Y_{3})
\end{split}
\end{equation}
where each new stabilizer is the summation of four Pauli operator. For the stabilizers in the bulk of the 1-D chain, there are two $\sw^\dag$ on them.
Take $X_{3}X_{4}$ and $-Z_{3}Z_{4}$ as example, $U_{2,3}$ and $U_{4,5}$ are performed on them. As a result, the corresponding stabilizers are:
\begin{equation}
  \begin{split}
  &(U_{2,3}\otimes U_{4,5}) X_{3}X_{4}(U_{2,3}\otimes U_{4,5})^{\dagger}=\\
 &(X_{2}I_{3}+I_{2}X_{3}+Y_{2}Z_{3}-Z_{2}Y_{3})\otimes\\
 &(X_{4}I_{5}+I_{4}X_{5}-Y_{4}Z_{5}+Z_{4}Y_{5}),\\
-&(U_{2,3}\otimes U_{4,5}) Z_{3}Z_{4}(U_{2,3}\otimes U_{4,5})^{\dagger}=\\
-&(Z_{2}I_{3}+I_{2}Z_{3}+Y_{2}X_{3}-X_{2}Y_{3})\otimes\\
&(Z_{4}I_{5}+I_{4}Z_{5}-Y_{4}X_{5}+X_{4}Y_{5}).
  \end{split}
\end{equation}
Both are the linear combinations of 16 Pauli tensors. For other stabilizers, similar forms could be obtained in this way.

In total, the number of these Pauli tensors is almost $32N$ which scales linearly with the qubit number. In fact, the measurement effort can be further reduced, as several Pauli tensors can be grouped and measured in one local measurement setting (LMS) simultaneously. For the Pauli tensors from $S_{2k-1}'=UX_{2k-1}X_{2k}U^{\dagger}$, we introduce the following expression for the example case $N=10$,

\begin{equation}
\begin{aligned}
&X_1\otimes(X_2X_3, Y_2Z_3, Z_2Y_3)\otimes (X_4X_5, Y_4Z_5, Z_4Y_5)\\
&\otimes (X_6X_7, Y_6Z_7, Z_56_7)\otimes(X_8X_9, Y_8Z_9, Z_8Y_9)\otimes X_{10},
\end{aligned}
\end{equation}
where one can periodically select one of the three Pauli tensors in every brace to construct one LMS, such that these LMSs cover all the possible Pauli operators from $S_{2k-1}'$. That is, select same tensor in (2,3) and (6,7) and the same Pauli operator in (4,5) and (8,9). Thus, only 9 LMSs are needed here. Following the same way, we can also find another 9 LMSs for the Pauli operators from $S_{2k}'=UZ_{2k-1}Z_{2k}U^{\dagger}$. As a result, totally only a constant number of $18$ LMSs is needed to obtain all the expectation values of the stabilizers.

To evaluate the robustness of our witness, we apply the white noise model $\rho_\mathrm{pre}=p\mathbb{I}/2^N+(1-p)\ket{\Psi}\bra{\Psi}$, and it shows that the noise tolerance is $p=\frac{3}{4N}$. As $N=10$, it equals $7.5\%$. 
In fact, by utilizing the trade-off between the robustness and the measure budget \cite{Guhne2007Toolbox,Qi2019Efficient}, one can enhance the noise tolerance of the witness further. In Ref.~\cite{Zhang2020generalized}, some of us further generalize the witness here by utilizing more local measurement settings, and the witnesses constructed there would be more suitable to realize in other quantum systems, such as superconducting-qubit. In particular, a 
numerical algorithm is developed in Ref.~\cite{Zhang2020generalized} to search for the witness with the optimal noise tolerance under given measurement settings. 
For instance, one can reach the noise tolerance $p=\frac{3}{16(1-2^{-N/2})}$ with about $2\cdot 3^{(N/2-1)}$ settings.

\subsection{Entanglement detection with homogeneous measurements}\label{Sec:homowitness}
The entanglement witness shown in Sec.~\ref{Sec:fidelity} based on the fidelity estimation can detect GME, with a few number of LMSs. However, in each of the LMS, the Pauli operators may be not the same, such as $X_1Y_2Z_3\cdots$. This kind of inhomogeoneus measurement needs additional local basis rotation, which is challenging for the current cold atom system.

In this section, we reduce the measurement efforts further by only considering the homogenous measurement, say $O^{\otimes N}$. In particular, we only need two LMSs, $X^{\otimes N}$ and $Z^{\otimes N}$. Note that $X^{\otimes N}$ and $Y^{\otimes N}$ are symmetric for our generation, thus we only need to measure one of them. Detailed illustration on this point is given in Method.

Instead of detecting GME, the protocol here aims to detect the full entanglement property defined in Eq.~\eqref{Eq:fully}, i.e., not separable with respective to any bipartition. First, let us show the following Lemma which plays an important role of the detection.
\begin{lemma}\label{Th:anti}
For a $k$-qubit quantum state $\rho$ with $k$ being even, if it is $\mathcal{P}_2$-separable, with $\mathcal{P}_2=\{A, \bar{A}\}$ and the number of qubits contained in $A$ and $\bar{A}$ being odd, the following inequality holds,
\begin{equation}\label{Eq:anti}
\begin{aligned}
|\langle X^{\otimes k}\rangle| +|\langle Y^{\otimes k} \rangle| +|\langle Z^{\otimes k}\rangle|\leq 1,
\end{aligned}
\end{equation}
where $\langle O\rangle=\mathrm{Tr}(\rho O)$ is the expectation value.
\end{lemma}

The proof of Lemma is based on the anticommutative relation \cite{Toth2005stabilizer,Huber2016anti} on subsystems $A$ and $\bar{A}$ respectively. As $k=2$, it becomes to the common criterion for the Bell state. As a result, the violation of the bound in Eq.~\eqref{Eq:anti} indicates that the underlying even-qubit state is non-separable regarding any odd-odd bipartitions.

Second, we give the following observation which shows the relation of the entanglement property of the state and its reduced density matrix (RDM).
\begin{observation}\label{Th:RDM}
Suppose a quantum state $\rho'_{B,\bar{B}}$ is created from $\rho_{A,\bar{A}}$ by local operation and classical communications(LOCC) $\Lambda$ with respective to the bipartiton $\{A,\bar{A}\}$, i.e.,
\begin{equation}\label{}
\begin{aligned}
\rho'_{B,\bar{B}}=\Lambda_{A\rightarrow B, \bar{A}\rightarrow \bar{B}}(\rho_{A,\bar{A}}),
\end{aligned}
\end{equation}
If $\rho_{A,\bar{A}}$ is separable for the bipartition $\{A, \bar{A}\}$, $\rho'_{B,\bar{B}}$ is also separable for the bipartition $\{B, \bar{B}\}$; in other words,
if $\rho'_{B,\bar{B}}$ is entangled for $\{B, \bar{B}\}$, $\rho_{A,\bar{A}}$ is also entangled for $\{A, \bar{A}\}$. 
\end{observation}
The observation holds as LOCC can not create entanglement. Specifically, the partial trace operation is an example of LOCC, where $B, \bar{B}$ are subsystems of $A, \bar{A}$ respectively.

By Definition \ref{Def:fully}, an $N$-qubit quantum state is fully entangled, if it cannot been written in the following form
\begin{equation}\label{Eq:weaksep}
\begin{aligned}
\rho\neq\sum_ip_i\rho_A^i\otimes \rho_{\bar{A}}^i,
\end{aligned}
\end{equation}
for all possible bipartitions $\{A, \bar{A}\}$ of the whole system. In the following, we employ Lemma \ref{Th:anti} and Observation \ref{Th:RDM} time and time again on RDMs of the 1-D prepared state, and finally certify its full entanglement. The RDM of the prepared state $\rho_\mathrm{pre}$, for example, on the first two qubits, is denoted by $\rho_{12}$. Note that the given expectation values are from the perfect target state. The practical value in the experiment may deviate from it, but can still reveal the entanglement property if the deviation is not too large. Detailed noise analysis are shown in Method.

To certify the full entanglement, we prove it by contradiction. Specifically, we first assume that the state can be written as right hand side of Eq.~\eqref{Eq:weaksep} for some $\{A, \bar{A}\}$, and then show that all the qubit are either in $A$ or in $\bar{A}$, which is actually not separable and leads to the contradiction.

Here we take $N=6$ for simplicity, and it can be directly generalized to any even qubit number $N$. The procedure is listed as follows.
\begin{enumerate}
  \item For the RDM $\rho_{12}$, one has
  \begin{equation}\label{Eq:2Vio}
\begin{aligned}
|\langle X_1X_2 \rangle| +|\langle Y_1Y_2 \rangle| +|\langle Z_1Z_2 \rangle|=\gamma_2>1,
\end{aligned}
\end{equation}
with $\gamma_2=1.5$ for the target state. It indicates that $\rho_{12}$ is non-separable on account of Lemma \ref{Th:anti}. Thus qubits $1,2$ should both be contained in $A$ or $\bar{A}$, otherwise it violates Observation \ref{Th:RDM}. Without loss of generality, we assume $1,2$ are in $A$. Similarly, $\rho_{13}$ also satisfies the inequality in Eq.~\eqref{Eq:2Vio}, and thus $1,2,3$ are all in $A$.
\item In fact, one can not proceed along the chain, since the RDMs $\rho_{23}, \rho_{34}\cdots$ are not entangled. Thus we consider correlations involving more qubits. For the RDM $\rho_{1234}$, one has
  \begin{equation}\label{Eq:4Vio}
\begin{aligned}
\left|\left\langle \bigotimes_{i=1}^4X_i\right \rangle\right| +\left|\left\langle \bigotimes_{i=1}^4Y_i\right \rangle\right| +\left|\left\langle \bigotimes_{i=1}^4Z_i\right \rangle\right|=\gamma_4>1,
\end{aligned}
\end{equation}
with $\gamma_4=1.5$ for the target state. It indicates that $\rho_{1234}$ is non-separable for the bipartition $\{123,4\}$ due to Lemma \ref{Th:anti}. Thus qubits $1,2,3,4$ should all be contained in $A$, due to Observation \ref{Th:RDM} and Step 1.
\item Same as Step 1, by applying Eq.~\eqref{Eq:2Vio} for $\rho_{46}$ and $\rho_{56}$, one can proceed to conclude that qubits $4,5,6$ should all be contained in $A$ or $\bar{A}$. Due to the result $\{1,2,3,4\}\subset A$ got in Step 2, we can arrive at the result that all the qubits are contained in $A$. This is contradict to the separable state on the right hand side of Eq.~\eqref{Eq:weaksep}. Thus the prepared state owns full entanglement.
\end{enumerate}

\subsection{Entanglement detection with reverse evolution}
Compared with the previous two entanglement detection protocols, the third one is more intuitional and aims to reveal GME. As shown in Sec.~\ref{Sec:fidelity}, it is not easy to obtain the fidelity to the target non-stabilizer state. The first method should use several non-homogenous LMSs to lower bound the fidelity.

In this section, we indirectly estimate the fidelity between the prepared state and the target state by evolving the entangling process backwards. Since the fidelity to both the Bell state and product state can be measured in the experiment, we can indirectly estimate the fidelity to $\ket{\Psi}$ and certify GME according to Lemma \ref{Th:Fwitness}. The reverse evolution follows the state preparation process shown in Sec.~\ref{subsec:Scheme}.
\begin{figure}[htbp]
  \centering
  \includegraphics[width=0.8\linewidth]{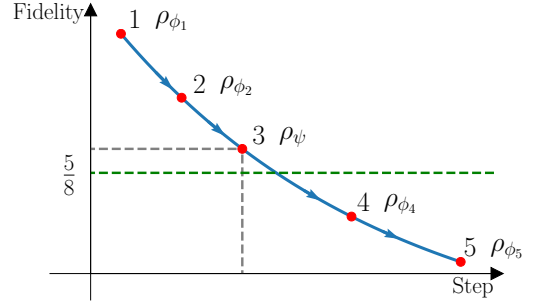}
  \caption{Illustration of reverse revolution. At point 3, the state preparation is accomplished; At point 4 and point 5, the system is under reverse evolution. In the perfect case, the state can return back to Bell pairs and Néel state. Here, due to the noise and decoherence, we denote the corresponding mixed state as $\rho_{\phi_i}$, and the fidelity is $\tr(\rho_{\phi_i}\Phi_i)$ to the perfect case. Note that one can estimate the fidelity to $\Phi_1,\Phi_5$ with one measurement settings $Z^{\otimes 10}$, and the one to $\Phi_2,\Phi_4$ with two settings $X^{\otimes 10},Z^{\otimes 10}$.}
  \label{fig:reverse}
\end{figure}
\begin{itemize}
\item Operate $\sqrt{\mathrm{SWAP}}$ gates parallelly on the qubit pairs $(2,3), (4,5),\cdots,(8,9)$.
\begin{equation}
\begin{aligned}
\ket{\Phi_4}=&\bigotimes_{k=1}^4\sqrt{\mathrm{SWAP}}_{\{2k,2k+1\}} \ket{\Psi}\\
=&\bigotimes_{k=1}^5\frac1{\sqrt{2}}(\ket{01}+\ket{10})_{\{2k-1,2k\}}.
\end{aligned}
\end{equation}
\item Adjust the relative phase between $\ket{01}$ and $\ket{10}$ in each of the Bell pairs in $\ket{\Phi_4}$.
\begin{equation}
\begin{aligned}
\ket{\Phi_4'}=U_{\mathrm{phase}}^\dag\ket{\Phi_4}=\bigotimes_{k=1}^5\frac1{\sqrt{2}}(\ket{10}+i\ket{01})_{\{2k-1,2k\}}.
\end{aligned}
\end{equation}
\item Operate $\sqrt{\mathrm{SWAP}}$ gates parallelly on the qubit pairs $(1,2), (3,4),\cdots,(9,10)$.
\begin{equation}
\begin{aligned}
\ket{\Phi_5}&=\bigotimes_{k=1}^5\sqrt{\mathrm{SWAP}}_{\{2k-1,2k\}} \ket{\Phi_4'}\\
&=\ket{1010101010}.
\end{aligned}
\end{equation}
\end{itemize}

In the above reverse protocol, the state is evolved backwards finally to the initial product state. Note that here we list the perfect reverse evolution, that is,  $\ket{\Phi_4}=\ket{\Phi_2}$ and $\ket{\Phi_5}=\ket{\Phi_1}$ shown in the generation protocol in Sec.~\ref{subsec:Scheme}.  The practical states actually deviate from them due to noises, denoted by $\rho_{\phi_i}$ for $i=1,2\cdots,5$, and the fidelity decreases due to the imperfection of the underlying gates. 

One can estimate the fidelity to product states $\ket{\Phi_1}, \ket{\Phi_5}$ with the $Z^{\otimes 10}$ measurement setting, entangled-pair states $\ket{\Phi_2'}, \ket{\Phi_4'}$ with $XYXY\cdots XY$ and $Z^{\otimes 10}$, and Bell-pair states $\ket{\Phi_2}, \ket{\Phi_4}$ with $X^{\otimes 10}$ and $Z^{\otimes 10}$. In Fig.~\ref{fig:reverse}, we show an illustration. By fitting the fidelity values from the experiment, one can indirectly estimate the fidelity of the prepared state $\rho_\mathrm{pre}$ to the target state $\ket{\Psi}$, which can be compared with the theoretical bound for GME, say $\frac{5}{8}$, denoted by the dotted horizon line in the figure. Note that $\sqrt{\mathrm{SWAP}}=\sqrt{\mathrm{SWAP}}^{\dag 3}$, i.e., it takes three times of evolution time, and one may needs to consider this point in the fitting. We remark that this entanglement detection method is intuitive, and one could make it more rigorous by assuming more specific noise models.

\section{Discussion}
In this work, we propose an experimental scheme to generate and characterize large-scale entanglement of cold atoms confined in the optical superlattice. The generation scheme utilizes the entangling gate induced by the superexchange interaction, and is robust to decoherence. To characterize the entanglement, we propose several complementary methods considering the experimental implementation feasibility. Moreover, it is straightforward to generate our scheme to the high dimension scenarios, and it is also interesting to construct some efficient entanglement verification tools for them. In summary, our entanglement generation and verification protocols are well tailored for the cold atom system, and lay the foundation for the further applications, such as measurement-based quantum computing. \comments{\red{Considering the recent development in MBQC with resource states generated by non-Clifford gates, it is interesting to apply the state here to study whether it can bring any advantage, say reduction of the single-qubit measurement to Pauli measurements.}}

\section{Methods}
\subsection{Proof of Proposition \ref{Th:Fwitness}}
\begin{proof}
To prove that Eq.~\eqref{Eq:Fwitness} is a legal GME witness, we need to show that the maximal fidelity between the separable state and our target state is upper bounded by $5/8$, that is
\begin{equation}
F_{\mathrm{max}}:=\max_{\rho_\mathrm{s}\in S_\mathrm{b}}\tr(\rho_\mathrm{s}\Psi)\leq 5/8.
\end{equation}
By the convexity of $S_\mathrm{b}$, we can reduce the maximization to the pure separable state
\begin{equation}
\begin{aligned}
F_{\mathrm{max}}=&\max_{\mathcal{P}_2}\max_{\Phi_\mathrm{s}\in S_\mathrm{b}^{\mathcal{P}_2}}\tr(\Phi_\mathrm{s}\Psi)
\\
=&\max_{\mathcal{P}_2} \lambda_1(\Psi,\mathcal{P}_2),
\end{aligned}
\end{equation}
where we maximize over $\Phi_\mathrm{s}$ from a given bipartition $\mathcal{P}_2$ and also all possible bipartitions. The second line is due to the fact that the maximization result over $\Phi_\mathrm{s}$ is just the square of the largest Schmidt coefficient of $\Psi$ with respective to $\mathcal{P}_2$ \cite{Bourennane2004Experimental},
where the Schmidt decomposition shows  $\ket{\Psi}=\sum_{i=1}^{d=\min{\{d_A,d_{\bar{A}}\}}}\sqrt{\lambda_i}\ket{\phi_i}_{A}\ket{\phi'_i}_{\bar{A}}$ and $\lambda_1\geq \lambda_2\geq\cdots \geq\lambda_d$.

In fact, we only need to consider the bipartitions where $A$ contains the first $n_A$ and $\bar{A}$ contains the last $n_{\bar{A}}=N-n_A$ qubits on the 1-D chain. That is, there is only one boundary between $A$ and $\bar{A}$. Other choices with more boundaries can lead to a smaller $\lambda_1$. Consider the one boundary scenario, there are two types of bipartations depending on that the boundary is between $\{2k-1,2k\}$ or $\{2k,2k+1\}$, as shown in Fig.~\ref{fig:split}.

\begin{figure}[htbp]
  \centering
  \includegraphics[width=0.9\linewidth]{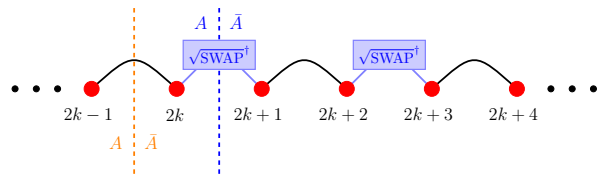}
  \caption{Bipartitions with different choices of boundary. Atom pairs connected with black line are Bell pairs. The dotted lines denote the boundary between $A$ and $\bar{A}$.}
  \label{fig:split}
\end{figure}

Based on the causality of the quantum gate and the fact that quantum gates inside the subsystem $A$ and $\bar{A}$ do not change the Schmidt number, we can determine $\lambda_1$ for both of cases.
For the $\{2k-1,2k\}$ case, $\lambda_1=\frac{1}{2}$, since we can ignore the latter $\sw^{\dag}$ and the entanglement spectrum is just the same with that of the Bell state. 

For the $\{2k,2k+1\}$ case, we just need to focus on the four-qubit state after the ignorance of the other distant $\sw^{\dag}$ gates, that is,
\begin{equation}\label{}
\begin{aligned}
\sqrt{\mathrm{SWAP}}^\dag_{\{2,3\}}\ket{\phi_{\mathrm{Bell}}}_{\{1,2\}}\otimes \ket{\phi_{\mathrm{Bell}}}_{\{3,4\}}.
\end{aligned}
\end{equation}
Here for denotation simplicity we take $k=1$ without loss of generality, and $\ket{\phi_{\mathrm{Bell}}}=\frac1{\sqrt{2}}\ket{01}+\ket{10}$. By calculation one finds that the reduced density matrix of the first two-qubit is
\begin{equation}\label{}
\begin{aligned}
\rho_{12}=\frac1{2}\phi_{\mathrm{Bell}}+\frac1{2}\mathbb{I}^{\otimes 2}/4,
\end{aligned}
\end{equation}
i.e., the mixture of the Bell state and the maximal mixed state. As a result, $\rho_{12}$ is a Bell-diagonal state and thus its four eigenvalues on the Bell basis is $\{\frac{5}{8}, \frac1{8}, \frac1{8}, \frac1{8}\}$ with 
$\lambda_1=\frac{5}{8}$. In summary, we prove that $F_{\mathrm{max}}=\frac{5}{8}$.
\end{proof}

\subsection{Proof of Proposition \ref{Th:lowerbound}}
\begin{proof}
We first show that $S_i'$ stabilizes the target $\ket{\Psi}$ by definition. Before the final layer of $\sw^{\dag}$ gates, the state is the product of a few of Bell pairs, denoted by $\ket{\Psi_{\mathrm{Bell}}}$, which is stabilized by $S_{i}$ defined in Eq.~\eqref{Eq:StaBell}. That is, $S_i\ket{\Psi_{\mathrm{Bell}}}=\ket{\Psi_{\mathrm{Bell}}}$. Thus, one has $S_i'\ket{\Psi}=US_iU^{\dag}U\ket{\Psi_{\mathrm{Bell}}}=U\ket{\Psi_{\mathrm{Bell}}}=\ket{\Psi}$.

Then we prove the inequality in Eq.~\eqref{Eq:OpIneq}. We remark that Eq.~\eqref{Eq:OpIneq} holds for any generalized stabilizer state. The $N$ independent stabilizers $S_i$ commute with each other and their common eigenstates together determine an orthogonal basis denoted by $\ket{\Psi_{\vec{\mathbf{x}}}}$. Here $\vec{\mathbf{x}}=x_1x_2\cdots x_N$ with each $x_i$ taking $1$ or $-1$ and $S_i'\ket{\Psi_{\vec{\mathbf{x}}}}=x_i\ket{\Psi_{\vec{\mathbf{x}}}}$. It is clear our target state $\ket{\Psi}=\ket{\Psi_{11\cdots1}}$. It is clear that both operators on the left and right of Eq.~\eqref{Eq:OpIneq} are diagonal in this basis, thus we can prove it by checking for the matrix elements for every $\Psi_{\vec{\mathbf{x}}}$. For $\vec{\mathbf{x}}=11\cdots1$, one has $1\geq \frac1{2}N-(N/2-1)=1$. For $\vec{x}$ which contains only one zero, such as $\vec{\mathbf{x}}=01\cdots1$, one has $0\geq \frac1{2}(N-1-1)-(N/2-1)=0$. For the $\vec{\mathbf{x}}$ containing more zeros, one can prove the inequality similarly.
\end{proof}

\subsection{Symmetry of the target state}
Remember that in principle we should apply three measurement settings $\{X^{\otimes N}, Y^{\otimes N}, Z^{\otimes N}\}$ to detect the full entanglement in Sec.~\ref{Sec:homowitness}. Here we show rigorously as follows that one does not need to worry about the direction on the $X-Y$ plane in every measurement. In fact, one only needs to measure $\langle X^{\otimes N} \rangle$ by the symmetry of the target state.

In the experiment, actually there is no reference pulse to decide the direction of the measurement, that is, one chooses a random angle $X_{\theta}=\cos(\theta) X+\sin(\theta) Y$ for the measurement. Note that there is a corresponding unitary (rotation around $Z$),
\begin{equation}\label{Eq:UrZ}
\begin{aligned}
U_{\theta}=\mathrm{e}^{-\mathrm{i}\frac{Z}{2}\theta}
\end{aligned}
\end{equation}
such that $X_{\theta}=U_{\theta} X U_{\theta}^{\dag}$.

As a result, suppose there is a standard $X$ direction, the actual measurement shows,
\begin{equation}\label{}
\begin{aligned}
\int_{\theta=0}^{2\uppi}\mathrm{Tr}(X_{\theta}^{\otimes N} \rho)d\theta&=\int_{\theta=0}^{2\uppi}\mathrm{Tr}(U_{\theta}^{\otimes N} X^{\otimes N}U_{\theta}^{\dag\otimes N} \rho)d\theta\\
&=\int_{\theta=0}^{2\uppi}\mathrm{Tr}(X ^{\otimes N} U_{\theta}^{\dag\otimes N} \rho U_{\theta}^{\otimes N})d\theta\\
&=\mathrm{Tr}\left(X^{\otimes N} \rho_{\mathrm{sym}} \right).
\end{aligned}
\end{equation}
where $\rho$ is the prepared state in the experiment, and $\rho_{\mathrm{sym}}\equiv\int_{\theta=0}^{2\uppi} U_{\theta}^{\dag \otimes N} \rho U_{\theta}^{\otimes N} d\theta$. 
In other words,  measuring the state $\rho$ in random direction is equivalent to ``twirling" the state, and the resulting state is symmetric with respective to the measurement direction. 
\begin{equation}\label{}
\begin{aligned}
\mathrm{Tr}(X_{\theta}^{\otimes N} \rho_{\mathrm{sym}} )=\mathrm{Tr}(X_{\theta^{'}}^{\otimes N} \rho_{\mathrm{sym}}).
\end{aligned}
\end{equation}
Specifically, $X=X_{\theta=0}$ and $Y=X_{\theta=\frac{\uppi}{2}}$, and one also has
\begin{equation}\label{}
\begin{aligned}
\mathrm{Tr}(Z^{\otimes N}  \rho_{\mathrm{sym}} )=\mathrm{Tr}(Z^{\otimes N} \rho),
\end{aligned}
\end{equation}
since $Z$ commutes with $U_{\theta}$ defined in Eq.~\eqref{Eq:UrZ}.

It is not hard to see that the entanglement of $\rho_{\mathrm{sym}}$ is not stronger than $\rho$, since $\rho_{\mathrm{sym}}$ is obtained from $\rho$ using LOCC, that is, the twirling can not increase the entanglement. Thus if one can detect the entanglement property of $\rho_{\mathrm{sym}}$, this property should also holds for $\rho$. 

On the other hand, the target state is a symmetric state on $X-Y$ plane, i.e., $U_{\theta}^{\otimes N} \Psi U_{\theta}^{\dag\otimes N}=\Psi, \forall \theta$. In fact, one can write the rotation unitary as 
\begin{equation}\label{}
\begin{aligned}
U_{\theta}^{\otimes N}=\mathrm{e}^{-\mathrm{i}\frac{\theta}{2}\left[\sum_{i=1}^N Z_i\right]}.
\end{aligned}
\end{equation}
Since our target state $\ket{\Psi}$ only has non-zero projection on the computational bases whose total spin is zero, i.e., half number of 0 and 1, the rotation unitary introduce the same phase for these bases, and the state is unchanged.

\subsection{Proof of Lemma \ref{Th:anti}}
\begin{proof}
Without loss of generality, we assume that the subsystem $A$ contains the first $k_1$ qubits, and $|A|=k_1$, $|\bar{A}|=k-k_1=k_2$ are both odd numbers. Since the left hand of Eq.~\eqref{Eq:anti} is a convex function of the state, thus we only need to consider the pure separable state in the form $\ket{\Psi_\mathrm{s}}=\ket{\Psi_{k_1}}\otimes \ket{\Psi_{k_2}}$, and the expectation value can be written apart as,
\begin{equation}\label{}
\begin{aligned}
& |\langle X^{\otimes k_1}\rangle||\langle X^{\otimes k_2}\rangle|  +|\langle Y^{\otimes k_1}\rangle||\langle Y^{\otimes k_2}\rangle|  +|\langle Z^{\otimes k_1}\rangle||\langle Z^{\otimes k_2}\rangle|\\
\leq &\sqrt{|\langle X^{\otimes k_1}\rangle|^2+|\langle Y^{\otimes k_1}\rangle|^2+|\langle Z^{\otimes k_1}\rangle|^2}\times\\
&\sqrt{|\langle X^{\otimes k_2}\rangle|^2+|\langle Y^{\otimes k_2}\rangle|^2+|\langle Z^{\otimes k_2}\rangle|^2},
\end{aligned}
\end{equation}
where the second line is due to Cauchy-Schwarz inequality. Since $k_1$ is an odd number, one can check that $X^{\otimes k_1}, Y^{\otimes k_1}$ and $Z^{\otimes k_1}$ anticommute with each other, and thus $|\langle X^{\otimes k_1}\rangle|^2+|\langle Y^{\otimes k_1}\rangle|^2+|\langle Z^{\otimes k_1}\rangle|^2\leq 1$ \cite{Toth2005stabilizer,Huber2016anti}. Similarly one has $|\langle X^{\otimes k_2}\rangle|^2+|\langle Y^{\otimes k_2}\rangle|^2+|\langle Z^{\otimes k_2}\rangle|^2\leq 1$ . As a result, we finish the proof.
\end{proof}
\subsection{Proof of Observation \ref{Th:RDM}}
\begin{proof}
The Observation is right by the definition of LOCC. Here we show the case for the partial trace operation by contradiction. Suppose $\rho_{A,\bar{A}}=\sum_ip_i\rho_A^i\otimes \rho_{\bar{A}}^i$ is separable. The partial trace $\tr_{A\rightarrow B}(\rho_A^i)=\rho_B^i$ and $\tr_{\bar{A}\rightarrow \bar{B}}(\rho_{\bar{A}}^i)=\rho_{\bar{B}}^i$, such that $\rho_{B,\bar{B}}=\sum_ip_i\rho_B^i\otimes \rho_{\bar{B}}^i$ is still separable, which contradicts with the assumption that it is entangled.
\end{proof}

\subsection{Influence from Operation Errors}
The fidelity of the state creation is affected by the error of each operation, caused by the decoherence effect and noises. For the large-scale entangled state, the degradation of fidelity and also the entanglement detection is more apparent compared to few-body states. In the meantime, the read-out error also affects the entanglement detection. As a result, it is meaningful to analyze the influence of various errors in the experimental implementation. 

Here we consider the errors in the following operations: the initial preparation of Néel state; the final projective measurement of spins; and the fidelity of the intermediate quantum gates. We show their influences on the entanglement detection of a ten-qubit system, i.e., $N=10$, by using the witness method in Sec.~\ref{Sec:homowitness}, which is more practical for experiments. 
As shown in the procedure of entanglement detection in Sec.~\ref{Sec:homowitness}, we should measure the witness value in Eq.~\eqref{Eq:anti} for a few of reduced density matrices of subsystems and also the whole system. Here we take the subsystem as 
\begin{equation}\label{ap:subsys}
\begin{aligned}
&\{1,2\},\{1,3\}; \{1,2,3,4\},\{1,2,3,5\};\\
&\{1-6\},\{1-5,7\}; \{1-8\},\{1-7,9\}
\end{aligned}
\end{equation}
which contain $n=2,4,6,8$ qubits.

\begin{figure}[htbp]
  \centering
  \includegraphics[width=0.9\linewidth]{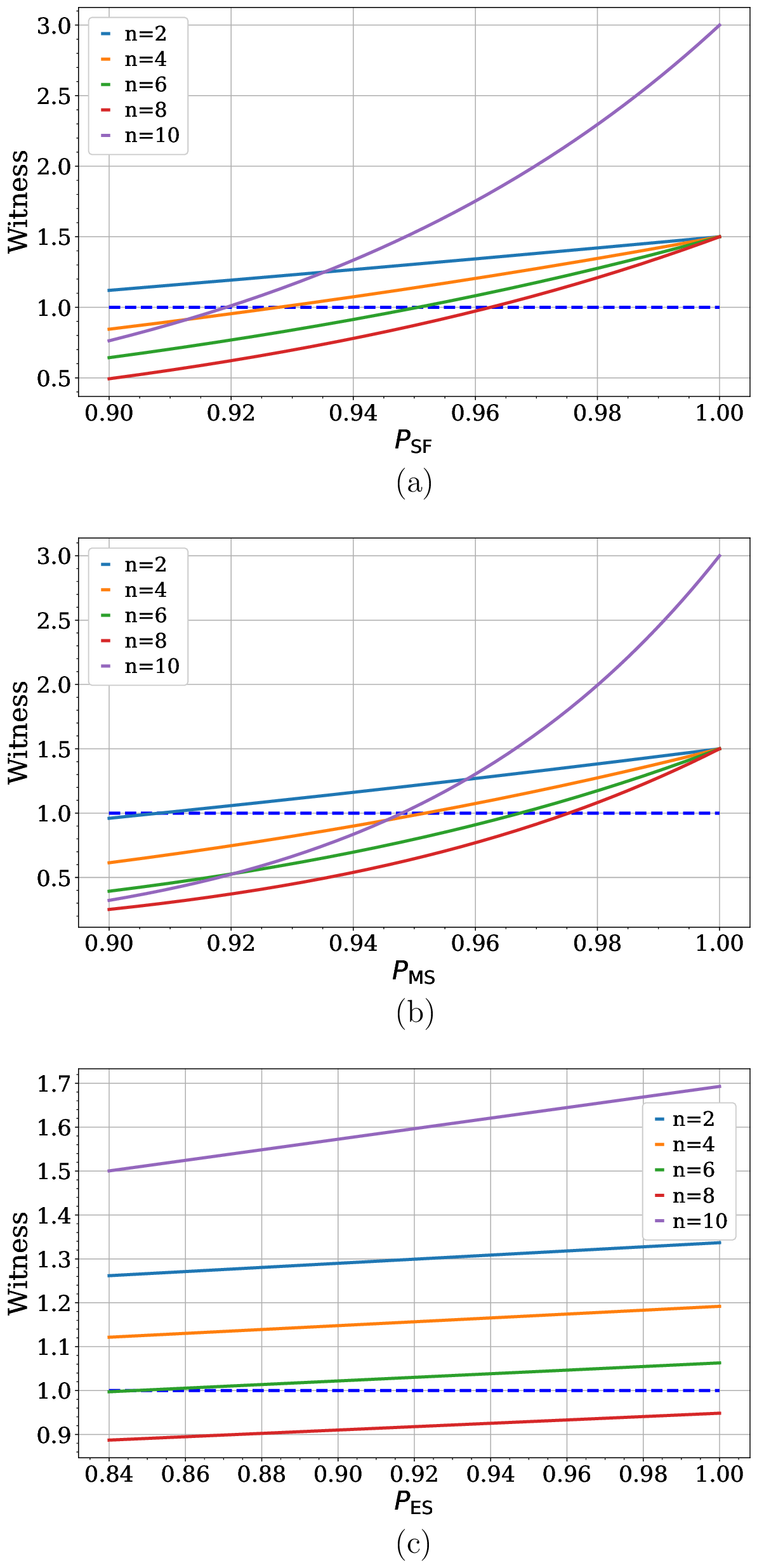}
  \caption{The value of witness of Eq.~\eqref{Eq:anti} in Sec.~\ref{Sec:homowitness} by taking account of errors in (a) spin-flip, (b) spin-measurement, and (c) entangling step. 
  In (c) $P_{\mathrm{SF}}$ and $P_{\mathrm{MS}}$ are taken to be 0.98 and 0.985, respectively. 
  Here the system contains $N=10$ qubits and the subsystems are chosen according Eq.~\eqref{ap:subsys} with qubit number $n=2,4,6,8,10$. The blue dashed line shows the lower bound of the witness that can certify the entanglement.}\label{Fig5}
\end{figure}

In the first step of Sec.~\ref{subsec:Scheme}, starting from the all-zero state $\ket{00\cdots0}$, one should flip the spins on the odd sites to prepare the Néel state in Eq.~\eqref{neel}. There is the probability of a non-flip for the odd site, and also the probability of a wrong flip for the even site. For simplicity, we assume both error probabilities are equal, and denote the probability of the correct flip as $P_{\mathrm{SF}}$ for each site. In this case, the system is initially prepared into a mixture of different product states weighted with corresponding probabilities. In Fig.~\ref{Fig5} (a), we plot the witness values of Eq.~\eqref{Eq:anti} for different subsystem size $n$ with respective to $P_{\mathrm{SF}}$. Similarly, in the final measurement, the probability to measure the spin of single atom correctly is denoted as $P_{\mathrm{MS}}$. That is, there is probability $1-P_{\mathrm{MS}}$ to recognize $\ket{0}$ as $\ket{1}$ or vice versa. The witness value with respective to $P_{\mathrm{MS}}$ is shown in Fig.~\ref{Fig5} (b). From these two figures, we found that the witness values are the same for the subsystem with the same qubit number, for example, $n=2$ case with $\rho_{1,2}$ and $\rho_{1,3}$. When there is no error, say $P_{\mathrm{SF}}=1$ and $P_{\mathrm{MS}}=1$,
all the witness values return $1.5$ for all the subsystems, except the whole system with the value $3$. One can see that even though there is some error, the witness can detect the entanglement as the value is larger than $1$. For larger subsystems, the values decay faster with $P_{\mathrm{SF}}$ and $P_{\mathrm{MS}}$. 

Similar as the six-qubit case discussed in Sec.~\ref{Sec:homowitness}, here for total system size $N=10$, one only needs to measure the witness for $\{\rho_{1,2},\rho_{1,3},\rho_{1,2,3,4},\rho_{1,2,3,5},\rho_{1,2,3,4,5,6},\\
\rho_{1,2,3,4,5,7},\rho_{7,8,9,10},\rho_{8,10},\rho_{9,10}\}$ to verify the full entanglement. As a result, the result of the subsystems with at most $n=6$ qubits decides the lower bound of the operation fidelity. According to Fig.~\ref{Fig5} (a) and (b), the bounds are 0.95 and 0.97 for $P_{\mathrm{SF}}$ and $P_{\mathrm{MS}}$ respectively.

At last, in Fig.~\ref{Fig5} (c), 
we study the influence of the fidelity of the entangling step between the above two steps by taking $P_{\mathrm{SF}}=0.98$ and $P_{\mathrm{MS}}=0.985$. Here we assume that this operation has a probability $P_{\mathrm{ES}}$ to be performed perfectly while it contributes zero to the witness with probability $1-P_{\mathrm{ES}}$, that is, essentially outputs a maximal mixed state. As shown in Fig.~\ref{Fig5} (c), the full entanglement could be verified when the fidelity of entangling step exceeds $0.85$.

\section{CODE AVAILABILITY}
The code that supports the findings of this study are available from the corresponding author upon reasonable request.

\section{DATA AVAILABILITY}
Data sharing is not applicable to this article as no data sets were generated or analyzed
during the current study.

\section{Acknowledgments}
This work was supported by the National Natural Science Foundation of China (NNSFC) No.~12125409, the Anhui Initiative in Quantum Information Technologies, and the Chinese Academy of Sciences. 
X.~Ma acknowledges the support by the NNSFC No.~11875173 and No.~12174216 and the National Key Research and Development Program of China No.~2019QY0702 and No.~2017YFA0303903.

\section{COMPETING INTERESTS}
The authors declare that there are no competing interests.

\section{AUTHOR CONTRIBUTIONS}
All authors contributed to the theory, numerical analysis, and writing up the manuscript. 
Y.~Z. and B.~X. contributed equally to this work.

\section{References}


%

\end{document}